\documentclass[rsi,graphicx,reprint]{revtex4-1}

\usepackage{booktabs}
\usepackage[utf8]{inputenc}
\usepackage[T1]{fontenc}
\usepackage{hyperref}
\usepackage{todonotes}
\usepackage{libertine}

\draft 

\begin{document}


\title{Fast Silicon Carbide MOSFET based high-voltage push-pull switch for charge state separation of highly charged ions with a Bradbury-Nielsen Gate} 



\author{Christoph Schweiger}
\email[]{christoph.schweiger@mpi-hd.mpg.de}
\affiliation{Max-Planck-Institut für Kernphysik, Saupfercheckweg 1, 69117 Heidelberg, Germany}

\author{Menno Door}
\affiliation{Max-Planck-Institut für Kernphysik, Saupfercheckweg 1, 69117 Heidelberg, Germany}

\author{Pavel Filianin}
\affiliation{Max-Planck-Institut für Kernphysik, Saupfercheckweg 1, 69117 Heidelberg, Germany}

\author{Jost Herkenhoff}
\affiliation{Max-Planck-Institut für Kernphysik, Saupfercheckweg 1, 69117 Heidelberg, Germany}

\author{Kathrin Kromer}
\affiliation{Max-Planck-Institut für Kernphysik, Saupfercheckweg 1, 69117 Heidelberg, Germany}

\author{Daniel Lange}
\affiliation{Max-Planck-Institut für Kernphysik, Saupfercheckweg 1, 69117 Heidelberg, Germany}

\author{Domenik Marschall}
\affiliation{Max-Planck-Institut für Kernphysik, Saupfercheckweg 1, 69117 Heidelberg, Germany}

\author{Alexander Rischka}
\affiliation{Max-Planck-Institut für Kernphysik, Saupfercheckweg 1, 69117 Heidelberg, Germany}

\author{Thomas Wagner}
\affiliation{Max-Planck-Institut für Kernphysik, Saupfercheckweg 1, 69117 Heidelberg, Germany}

\author{Sergey Eliseev}
\affiliation{Max-Planck-Institut für Kernphysik, Saupfercheckweg 1, 69117 Heidelberg, Germany}

\author{Klaus Blaum}
\affiliation{Max-Planck-Institut für Kernphysik, Saupfercheckweg 1, 69117 Heidelberg, Germany}


\date{\today}

\begin{abstract}
In this paper we report on the development of a fast high-voltage switch, which is based on two enhancement mode N-channel Silicon Carbide Metal Oxide Semiconductor Field-Effect Transistors in push-pull configuration.
The switch is capable of switching high voltages up to 600 V on capacitive loads with rise and fall times on the order of 10 ns and pulse widths $\geq$20 ns.
Using this switch it was demonstrated that from the charge state distribution of bunches of highly charged ions ejected from an electron beam ion trap with a specific kinetic energy, single charge states can be separated by fast switching of the high voltage applied to a Bradbury-Nielsen Gate with a resolving power of about 100.
\end{abstract}


\maketitle 

\section{Introduction}\label{sec:intro}

Many experiments addressing fundamental physics require highly charged ions (HCI), see e.g.~\cite{Sturm19, Micke20, Stark21, Filianin21}, which are produced in ion sources employing different ionization mechanisms, e.g. electron impact ionization such as in electron beam ion traps (EBITs)~\cite{Levine88}.
In EBITs, a distribution of charge states is produced of which often only a single charge state is experimentally required.
In order to separate the charge state of interest and remove the background of unwanted species, different charge-to-mass ratio selective techniques can be employed such as Wien-type velocity filters, sector magnets or Time-of-Flight (ToF) separation~\cite{Wolf12}.
Here, resolving powers of 20-200 are achievable, using a Wien-type velocity filter~\cite{Schmidt09} (depending on the aperture) and around 150 for the sector magnet~\cite{Rao99}.
For the single-pass ToF separation reported here a resolving power of around 100 is possible.

In our specific setup, a compact, room-temperature EBIT~\cite{Micke18, Schweiger19} is used for the production of HCI, which are extracted and transported to a Penning-trap setup for high-precision mass spectrometry~\cite{Repp12}.
In the Penning trap, only a single HCI is stored, requiring background reduction and the selection of a single charge state.
Upon extraction from the EBIT, a bunch of ions is accelerated by an electrostatic potential. 
This results in slightly different velocities of the ions depending on their charge state $v \sim \sqrt{q}$ (assuming the same mass).
The ions in higher charge states propagate slightly faster through the beamline than the ones in lower charge states and therefore arrive earlier at the detector plane.
Individual charge states can now be selected by deflecting all other species, e.g.~by applying a voltage to some electrode and switching it to ground only for the short time window when the charge state of interest passes the electrode.
This concept is experimentally realized using a Bradbury-Nielsen Gate (BNG)~\cite{Bradbury36, Yoon07, Wolf12, Brunner12} combined with a fast switching electronic circuit in order to resolve individual charge states.

Fast and efficient switching of high voltages is today commonly used in power electronics, e.g. in power supplies and driving electronics for electrically powered vehicles.
Due to the rising number of required devices, different types of power Metal Oxide Semiconductor Field-Effect Transistors (MOSFETs) for these applications with very fast rise and fall times as well as low drain to source resistances are developed and commercially available today.
In the context of this paper, two power MOSFETs are used in order to build a fast switching electronic circuit controlled using a~$5\,\mathrm{V}$ TTL-logic signal.
The application requires that the duration of the high-voltage pulse can be as short as $20\,\mathrm{ns}$ and that the time it takes to switch from one voltage to the other is on a timescale of around $10\,\mathrm{ns}$ for voltages of up to $500\,\mathrm{V}$.
Similar solid-state switches suitable for even significantly higher voltages are commercially available today and are summarized in Table \ref{tab:HVsolidstateswitches}.
\begin{table*}[tbp]
	\begin{tabular}{lllll}
		\hline
			Company & Switch type & rise/fall times (ns) & pulse width (ns) & Voltage rating (V) \\
		\hline
			Behlke Power Electronics GmbH & HTS61-05~\cite{Behlke} & 5 & 50.0 & 6000 \\
			CGC Instruments & NIM-AMX500-3~\cite{cgcinstr} & 20 & 150 & 500\\
			Berkeley Nucleonics Corp. & Model 6040-310H~\cite{berkeley} & 15 & 25 & 800 \\
			Berkeley Nucleonics Corp. & Model 6040-202H~\cite{berkeley} & 5 & 12 & 300 \\
		\hline
	\end{tabular}
	\caption{Comparison of commercially available solid-state high-voltage switches.}
	\label{tab:HVsolidstateswitches}
\end{table*}
Except for the two electrical modules for the switch from Berkeley Nucleonics, these solid state switches do not fulfill all our requirements for the switching, e.g. the minimal pulse width.
Furthermore, when switching high-voltage, the load (e.g.~a capacitive load) and the signal path from the switching device to the electrodes has to be considered.
For the commercial devices the connections are typically done using either BNC or SHV (secure high-voltage) connector and the respective cables.
Depending on the size of the device, these have to be installed in a dedicated place, sometimes even with a dedicated power supply which further increases the cable length and can potentially lead to a performance reduction due to the increased capacitance.
The switching circuit presented in this paper, specifically also the printed circuit board which hosts the switching circuit can be easily adapted to the experimental requirements by the builder and in this way generally reduces cable lengths and allows a better control of the signal paths' impedance.
Ultimately it can be even designed to be placed within the vacuum directly attached to the relevant electrode.

In addition to the commercial devices listed above, there are a number of publications of high-voltage solid-state switches with similar performances but different designs optimized for the specific application and not necessarily using the push-pull principle~\cite{Baker90, Continetti92, Dedman01a, Dedman01b, Feng11, Azizi20}.
This can be favorable e.g.~for the use as Pockels cell drivers in Q-switched lasers, where a very fast rise-time is relevant but the pulse duration and the fall-time are not similarly important~\cite{Rutten07}.

In the following sections, the fast switching electronic circuit is described and measurements of the switching performance are shown.
Finally the separation of individual charge states is experimentally demonstrated, showing a resolving power on the order of~100.

\section{Electronic circuit}\label{sec:schematic}

\begin{figure*}
\centering
	\includegraphics[width=0.8\textwidth]{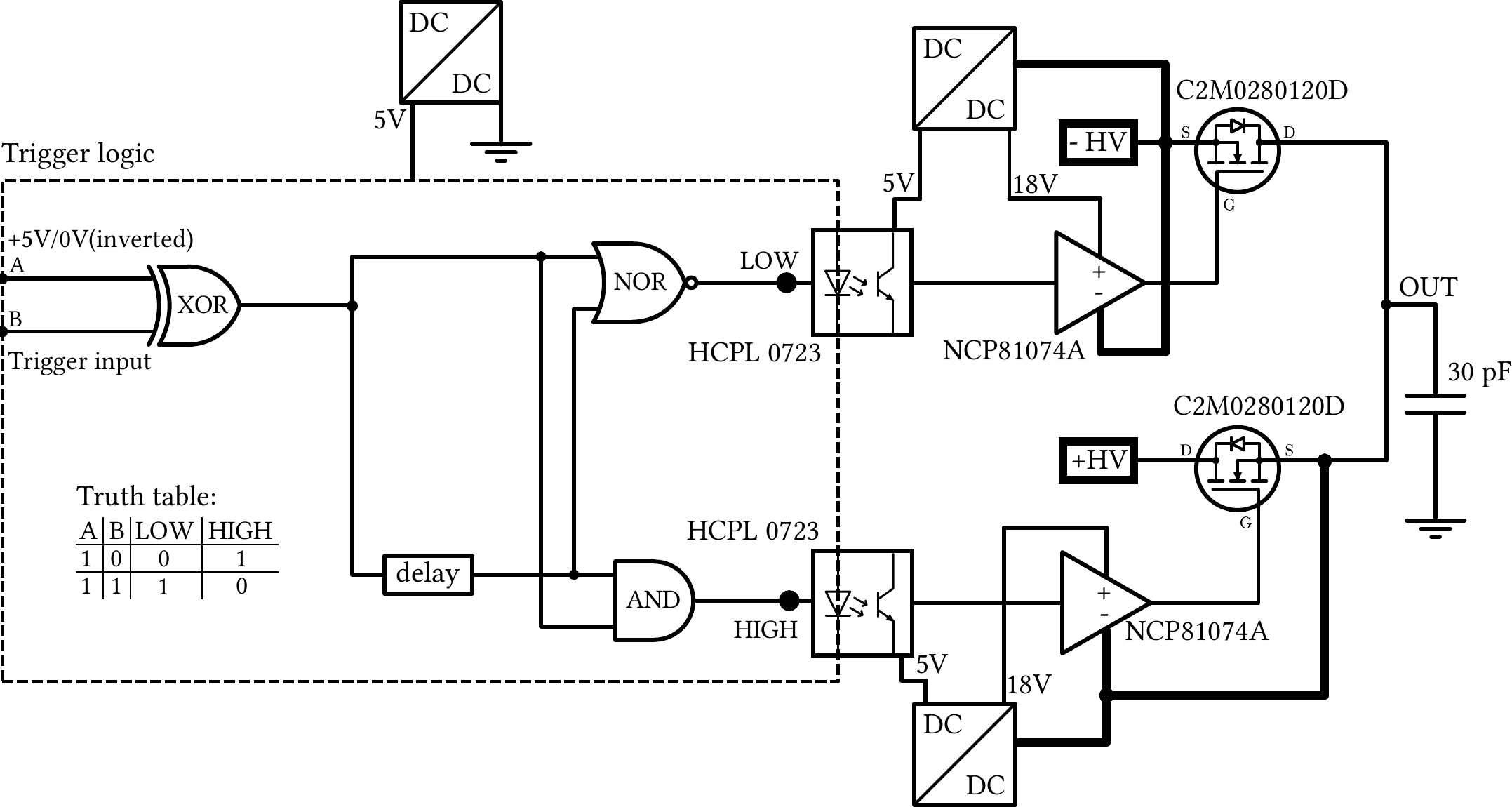}%
	\caption{Schematic overview of the electronic circuit. 
	The dashed box surrounds the logic gates that form the two trigger pulses for the gate drivers from a single input trigger pulse. 
	For the two logic gate combinations a combined truth table is shown for the non-inverted case, i.e. when A is ''high``, such that the output depends only on the trigger input B.
	The input A can be also set to $0\,\mathrm{V}$ which inverts the output of the trigger logic.
	The results given in the truth table correspond to the points ''LOW`` and ''HIGH``.
	Following the trigger logic, an optical insulator separates the floating grounds of the gate drivers from the trigger ground.
	In the final stage, the optically insulated trigger signal reaches the gate drivers, which produce an $18\,\mathrm{V}$ signal in order to quickly charge the gate capacitance of the MOSFET.
	A $30\,\mathrm{pF}$ capacitive load is shown on the output, which was used for laboratory testing of the switch and is comparable to the capacitive load of the BNG.}
	\label{fig:schematic}
\end{figure*}

A schematic overview of the electronic circuit of the push-pull switch is given in Figure~\ref{fig:schematic} showing the main electrical components.
The complete and detailed circuit can be found in the supplementary material.
The timing and pulse duration of the high-voltage pulse is controlled with a $5\,\mathrm{V}$ TTL trigger input.
The trigger logic is designed such that as long as the trigger signal is high (low), the higher (lower) voltage is active on the output.
In this way the pulse length of the high-voltage output pulse is controlled by the width of the trigger pulse $\Delta t_{\mathrm{TRIG}}$.
The trigger section consists of the logic gates, a delay timing element and the input side of the optocouplers (dashed box in Figure~\ref{fig:schematic}), which are all supplied from a single DC/DC converter with the ground connected to the main ground.
Each of the two trigger signals then passes through an optocoupler (Broadcom HCPL-0723~\cite{BroadcomOptocoupler}) which insulates the trigger logic part from the gate drivers and the MOSFETs.
In order to be able to switch voltages up to $\pm 500\,\mathrm{V}$ with rise and fall times in the few tens of nanosecond range, Silicon Carbide based MOSFETs (SiC-MOSFETs) have shown the best performance in our tests.
More specific, we used two N-channel enhancement mode SiC-MOSFETs (Wolfspeed CREE C2M0280120D~\cite{WolfspeedSiCMOSFET}), which were chosen specifically for their fast rise and fall times of around $10\,\mathrm{ns}$ and the approximately symmetric time constants.
This type of MOSFET has shown the best performance for our purposes compared to other MOSFETs from different manufacturers which were tested based on silicon, SiC and gallium nitride (GaN).
These MOSFETs were found by searching specifically for high-voltage and fast rise- and fall times which have roughly the same time constant.
Each of the MOSFETs had to be tested individually since the specifications listed in the datasheets are usually measured in specific conditions which do not necessarily reflect the ones that are used for the switch operation, e.g. the load on the output as well as the used gate driver are mostly different.
Among the tested models were MOSFETs from ST~Microelectronics N.V., e.g.~STP45N40DM2AG~\cite{stMOSFET}, Texas Instruments~Inc., e.g.~LMG3410R150~\cite{TexasInstrumentsMOSFET} and GaN Systems~Inc., e.g.~GS-065-004-1-L~\cite{GaNSysMOSFET}.
Several more products from manufacturers such as Infineon Technologies AG, Microchip Technology Inc. (SiC-type) and Navitas Semiconductor (GaN-type) were also considered.
As newer MOSFET types with possibly faster characteristics become available, the electronic circuit can be easily adapted to use these together with an appropriate gate driver.
The two MOSFETs used in the circuit are controlled each by an individual gate driver (Onsemi~NCP81074A~\cite{OnsemiGateDriver}), which charges and discharges the MOSFET gate with rise and fall times as well on the order of~$10\,\mathrm{ns}$.
\begin{figure}[htbp]
	\includegraphics[width=\columnwidth]{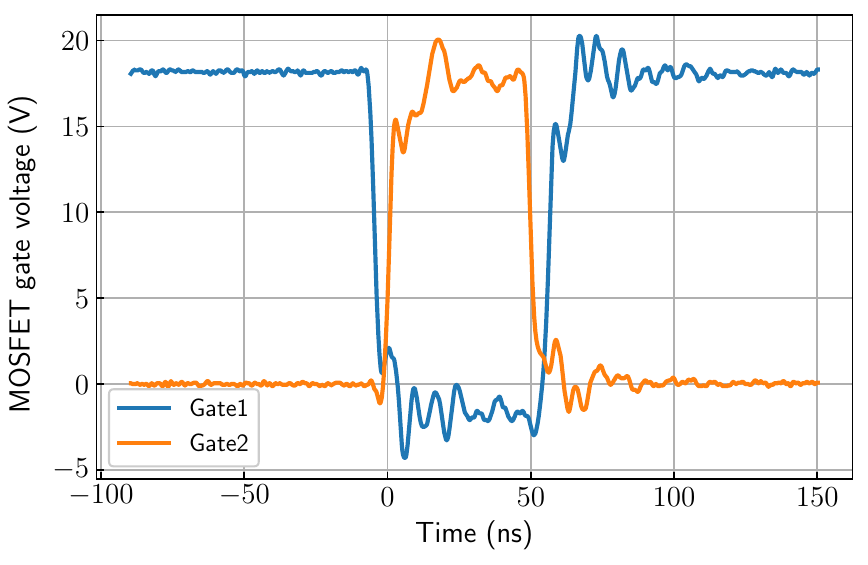}%
	\caption{MOSFET gate pulses for $\Delta t = 50 \,\mathrm{ns}$ pulse width. For details see text.}
	\label{fig:gatepulses}
\end{figure}

The measured pulses from the gate drivers are shown in Figure~\ref{fig:gatepulses}.
Clearly visible here is the switching sequence from one voltage to the other:
First, the MOSFET which is initially conducting (blue curve) is switched off before the other one (orange curve) is switched to the conducting mode.
When switching back to the original voltage, the reverse order is followed.
Due to the high voltage which is switched, the voltage on the source pin of the MOSFET will quickly become larger than the voltage on the gate, in which case the MOSFET is no longer conducting.
In order to avoid that the MOSFET switches off, the gate and the corresponding driver have to be floating together with the source voltages of the MOSFETs as indicated in Figure~\ref{fig:schematic} with the wider connection lines.
This is implemented by having a floating ground and an insulated DC/DC-converter for each of the gate drivers and the corresponding MOSFETs, which is insulated towards the trigger logic part with the optocouplers and connected to the source on the MOSFET. 
For the connected high voltages it is important that the +HV is always more positive than the -HV. In turn this means that -HV does not necessarily have to be negative.
In the printed circuit board layout of the switch the ground planes were clearly separated to avoid any coupling from the floating grounds to the main ground at high voltages.
Furthermore, in order to achieve fast rise and fall times, the connections between the MOSFETs and the capacitive load should be designed with a surface as large as possible in order to maximize the conduction on the surface (skin effect) and reduce parasitic inductance.

\section{Push-pull switching performance}\label{sec:switchperformance}

Initial testing of the electrical circuit was done in a bench setup using a similar capacitance as the BNG, which is on the order of 30 pF.
The capacitor is connected between the output of the switch and the main ground as shown in Figure \ref{fig:schematic}.
This setup is not fully realistic since the BNG is not connected to the main ground with one side but to a second switch.
In order to measure the rise and fall times ($t_{\mathrm{r}}$ and $t_{\mathrm{f}}$, respectively) as well as the pulse width $\Delta t$ of the high-voltage pulse on the capacitive load, a fast high-voltage probe~\cite{HVprobe} was used together with an oscilloscope~\cite{RSOszi} for data acquisition and measurement.

\begin{figure}[htbp]
	\centering
		\includegraphics[width=1.00\columnwidth]{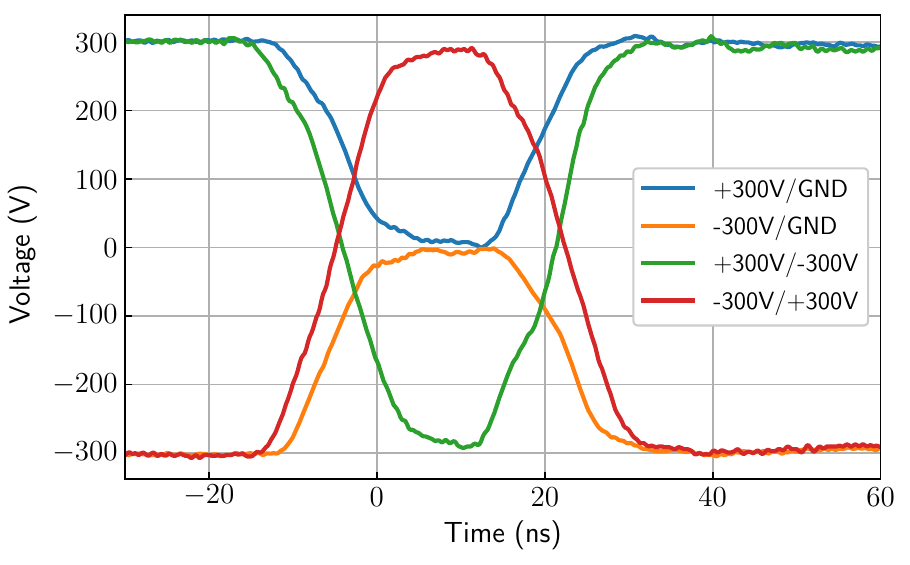}%
	\caption{Measured output voltage using the bench test setup with 30 pF capacitance connected to the output. 
	The rise and fall times as well as the measured pulse widths are listed in Table \ref{tab:switchingtimes}.
	Typical in our experiment~\cite{Filianin21} is the operation of switching a positive or a negative voltage to ground (blue and orange curves), see Section~\ref{sec:qsep}.
	Furthermore interesting is switching a positive to a negative voltage or vice versa (green and red curves).
	For all curves the trigger point is at $t = 0\,\mathrm{ns}$.}
	\label{fig:switchcurves}
\end{figure}

\begin{table}[hbp]
	\begin{tabular}{lllll}
		\hline
			switched voltage (V) & $t_r$ & $t_f$ & $\Delta t_{\mathrm{FWHM}}$ & $\Delta t_{\mathrm{TRIG}}$ \\
		\hline
			$+300 \rightarrow$ 0 & 9.4 & 11.2 & 23.2 & 20 \\
			$-300 \rightarrow$ 0 & 9.2 & 11.0 & 28.6 & 20 \\
			$+300 \rightarrow$ $-$300 & 12.4 & 13.8 & 26.4 & 20 \\
			$-300 \rightarrow$ $+$300 & 11.6 & 14.0 & 28.0 & 20 \\
		\hline
	\end{tabular}
	\caption{Measured rise- and fall times for different switch operations in ns for the switching curves shown in Figure~\ref{fig:switchcurves}. 
	The measured pulse width $\Delta t_{\mathrm{FWHM}}$ is slightly larger than the pulse width of the trigger pulse used for controlling the switch $\Delta t_{\mathrm{TRIG}}$.}
	\label{tab:switchingtimes}
\end{table}

The output voltage of the switch obtained with the described test setup are shown in Figure \ref{fig:switchcurves}.
Rise and fall times (10\% to 90 \%) and the pulse width (full width at half maximum - FWHM) measured with the built-in measurement function of the oscilloscope for ten averages are listed in Table \ref{tab:switchingtimes}.
The types of switch operations are illustrated for which the switch is typically used in our laboratory, that is switching a positive or negative voltage to ground potential (blue and orange curves) or switching from a positive to a negative voltage (green and red curves).
It is furthermore possible to switch from one voltage to another voltage of the same polarity which is not illustrated.
Figure \ref{fig:switchcurves} shows the switch curves for the shortest possible pulse width of 20 ns. For longer pulse widths the switching curves become significantly more ''rectangular``-shaped.
Compared to the commercial devices (see Table \ref{tab:HVsolidstateswitches}), the rise- and fall times as well as the pulse widths are on a similar level if only the numbers are compared.
Especially for the very fast models from Berkeley Nucleonics Corp. a thorough comparison with a similar load and in identical conditions would be interesting, since, as we have noted before, also the connections and cables which are used can change the performance.

\section{Charge-state separation of highly charged ions using a Bradbury-Nielsen gate}\label{sec:qsep}

\begin{figure*}
	\includegraphics[width=\textwidth]{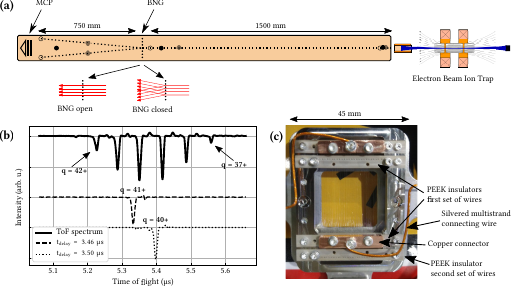}%
	\caption{An overview of the experimental setup is shown in the upper part of the Figure~(a).
	Following the ejection of HCI from an EBIT, the HCI with a kinetic energy of $4\,\mathrm{keV/q}$ propagate through the beamline and eventually pass the BNG before reaching an MCP detector at the end of the beamline.
	The black, grey and unfilled circles represent ions with different charge states. 
	In this illustration, the BNG is only open during the passage time of the ion charge state marked as black circles, leading to a deflection of the ion charge states in grey and unfilled circles.
	In (b), a measured ToF spectrum of ions is shown where individual charge states of highly charged $^{163}\mathrm{Dy}$ ions are detected on the MCP detector at different times following the ejection from the EBIT and can be observed as dips in the signal extracted from the anode behind the MCP detector plates.
	The two lower curves in (b) show the ToF spectrum of two individual charge states for which the voltage of the BNG wires was set to $\pm500\,\mathrm{V}$ and only switched to ground potential for a short period of $50\,\mathrm{ns}$ with two different time delay settings $t_{\mathrm{delay}}$ of the BNG trigger pulse.
	(c)~Shows a photograph of the BNG setup built into the experiment.
	The square-shaped aperture with the wires is 25 x 25$\,$mm in size.
	For details see text.}
	\label{fig:EBITsetup}
\end{figure*}

In our experimental setup~\cite{Filianin21}, bunches of highly charged ions are produced in an electron beam ion trap (EBIT)~\cite{Schweiger19, Micke18} and are extracted in bunches containing a distribution of charge states.
The developed switch is designed to supply the voltages for a BNG~\cite{Bradbury36} for the separation of individual charge states.
Upon extraction from the EBIT, the ion bunch is accelerated by a voltage $U = 4000\,\mathrm{V}$ and propagates through several ion optical elements before passing through the BNG and finally impinging on a microchannel plate detector (MCP) in Chevron configuration.
The final velocity of the ions, following an acceleration by a voltage $U$, differs slightly depending on the charge state $q_{\mathrm{ion}} = n \cdot e$ of the ions. Here, $n$ is the number of missing electrons and $e$ the elementary charge.
%
%
Thus, after propagating a finite distance $s$, the ions arrive at the MCP detector at slightly different times depending on their charge state:
\begin{equation}
t(q_{\mathrm{ion}}) = \sqrt{\frac{s^2 m}{2eUq_{\mathrm{ion}}}}.
\end{equation}
This assumes that only one ion species with mass $m$ is present in the EBIT.
A ToF spectrum of the ejected ion bunch arriving on the MCP detector is shown in Figure~\ref{fig:EBITsetup}~(b)~(solid curve), where a clear signal appears for each arriving charge state (from $q = 37+$ to $42+$) separated by about $70\,\mathrm{ns}$ from each other.
In order to separate the individual charge states, a BNG can be used in a similar way as in multi-reflection ToF mass spectrometry~\cite{Wolf12, Brunner12}.
A BNG consists of two sets of wires which are alternately arranged in parallel as shown in the photograph in Figure~\ref{fig:EBITsetup}~(c).
The wires, made from $60\,$µm stainless steel wire are individually wound through the PEEK insulator material and have a spacing of $0.5\,\mathrm{mm}$.
One set of wires is attached to the inner two PEEK insulators, the second set to the outer ones, such that the wires are alternately arranged in parallel without electrical contact.
In operation, one set of wires is set to a positive voltage while the other is set to a negative voltage which leads to a deflection of charged particles passing through the plane of the wires (''BNG closed`` in Fig.~\ref{fig:EBITsetup}~(a)). 
If there is no voltage applied, the ions can pass the BNG without being deflected (''BNG open``).
The general idea is to switch between these two states and thereby only allow the charge state of interest to pass the BNG, controlled by the precisely controlled timing of the switching process. 
In order to change the applied voltages very fast, the switch presented in Section~\ref{sec:schematic} is used.
Each set of wires is supplied from a separate switch, one switching from a positive voltage to ground, the other one from a negative voltage.
The connection is done via a $4\,\mathrm{mm}$ copper rod vacuum feedthrough and silvered multistrand wires on the vacuum side in order to maximize the conducting surface.
From the multistrand wire to the BNG wire, a solid block of copper is used which connects to all of the wires, see Figure \ref{fig:EBITsetup} (c).

The lower part of Figure \ref{fig:EBITsetup} (b) shows the ToF spectra for two different time delay settings $t_{\mathrm{delay}}$ of the trigger pulse ($50\,\mathrm{ns}$ pulsewidth) controlling the switching circuit.
This time delay is adjusted such that the BNG wires are switched from $\pm 500\,\mathrm{V}$ to ground potential when the selected charge state arrives at the position of the BNG wires, $1.5\,\mathrm{m}$ from the ejection point (around~$3.5\,\mu\mathrm{s}$ following the ejection).
The ions finally arrive at the MCP detector plates following another $75\,\mathrm{cm}$ of flight path after a total ToF of $\sim 5.4\,\mu\mathrm{s}$.
With this, the capability of separating individual charge states following each other in close succession in the ToF spectrum is demonstrated.
The two separated charge states arrive slightly delayed compared to the reference ToF spectrum.
Unlike the ideal model of a BNG, the real BNG can have residual voltages on the wires and the fast switching of the voltages on the BNG wires can cause an accelerating or decelerating electric fields when the ions are still close the plane of the wires.
This additional deceleration causes the ions to arrive delayed on the MCP detector and can be minimized by tuning the pulse length of the high-voltage pulse on the BNG wires and the time delay of the trigger pulse.

With the pulse width of $\Delta t_{\mathrm{FWHM}} = 50\,\mathrm{ns}$ and the ToF of about $t_{\mathrm{ToF}} = 5\,\mu\mathrm{s}$ we can estimate a resolving power of $R = \frac{t_{\mathrm{ToF}}}{\Delta t_{\mathrm{FWHM}}} \simeq 100$.
This is in the range where also other charge-to-mass ratio selective devices operate as mentioned in Section~\ref{sec:intro}.
A sufficient separation is already achieved at voltages around $\pm 200\,\mathrm{V}$ applied to the wires.
In an upgraded version of our experimental setup, a set of slits is used in order to cut away the deflected ions that still reach the MCP surface at lower voltages.
In our experimental setup, this fast switching circuit in combination with the BNG allowed the clear separation of individual charge states of HCI within a relatively short beamline of just a few meters using only moderate deflection voltages.

\section{Conclusion}\label{sec:conclusion}

In this paper a fast high-voltage push-pull switch with switching times on the order of 10 ns and short pulse widths of 20 ns is presented.
Results of the switch being used as a high-voltage switch for a BNG have been shown as well as the successful separation of highly charged ions following a ToF separation.
Due to its fast switching times the switch is meanwhile used in other experiments, e.g. for fast switching of piezo valves or the electrode potentials on Penning traps at SHIPTRAP~\cite{Ketelaer08}.

\section*{Supplementary material}
The detailed electrical circuit of the switch is available in the supplementary material.

\begin{acknowledgments}
This project was funded by the European Research Council (ERC) under the European Union’s Horizon 2020 research and innovation programme under Grant Agreement No. 832848-FunI and by the Deutsche Forschungsgemeinschaft under Grant number 273811115-SFB1225-ISOQUANT. 
Furthermore, we acknowledge funding and support by the Max-Planck-Gesellschaft and the International Max-Planck research school for precision tests of fundamental symmetries (IMPRS-PTFS). 
This work comprises parts of the Ph.D. thesis work of C.S. to be submitted to Heidelberg University, Germany. 
\end{acknowledgments}

\section*{Data Availability}
The data that support the findings of this study are available from the corresponding author upon reasonable request.

\bibliography{BNGswitch.bib}

\end{document}